\documentclass[sn-mathphys-ay]{sn-jnl}

\usepackage{graphicx}%
\usepackage{multirow}%
\usepackage{amsmath,amssymb,amsfonts}%
\usepackage{amsthm}%
\usepackage{mathrsfs}%
\usepackage[title]{appendix}%
\usepackage{xcolor}%
\usepackage{textcomp}%
\usepackage{manyfoot}%
\usepackage{booktabs}%
\usepackage{listings}%
\usepackage{array}%
\usepackage[linesnumbered,ruled,vlined]{algorithm2e}

\begin{document}

\title{Large scale study of primary school student performance relative to their LMS activity and socioeconomic demographics using a Bayesian Additive Regression Trees containing random effects}


\author*[1]{\fnm{Natalia} \sur{da Silva}}\email{natalia.dasilva@fcea.edu.uy }

\author[1]{\fnm{Bruno} \sur{Tancredi}}\email{brunotancredi00@gmail.com}
\author[1]{\fnm{Ignacio} \sur{Alvarez-Castro}}\email{ignacio.alvarez@fcea.edu.uy}
\equalcont{These authors contributed equally to this work.}


\affil[1]{\orgdiv{Instituto de Estadística}, \orgname{Facultad de Ciencias Económicas y de Administración - Universidad de la República}, \orgaddress{\street{Gonzalo Ramirez 1926}, \city{Montevideo}, \postcode{11200}, \state{Montevideo}, \country{Uruguay}}}



\abstract{
Using  data collected on almost every 9-12 years old student in Uruguay, we show how to apply Bayesian Additive Regression Trees (BART) with random effects to study performance association with Learning Managment System (LMS) activity and socioeconomic status. Performance data is joined with LMS activity pattern data. BART is chosen because it is possible to include school-level random effects. The model can be used for early identification of at-risk students, and highlights schools that are successful or need intervention. An interesting finding is that high levels of LMS usage show larger positive effects on performance in low socioeconomic status.
}



\maketitle

\section{Introduction} \label{sec:Intro}
Since 2007, Uruguay has implemented the One Laptop per Child (OLPC) program in primary education, locally known as Plan CEIBAL (\url{https://ceibal.edu.uy}). This initiative serves as a model for developing socio-educational programs that integrate one-to-one technology with human development, digital inclusion, and equal opportunities. Initially implemented in public primary schools, the program was extended to secondary schools in 2011 and played a key role during the COVID-19 pandemic. CEIBAL allows us to have usage data from students and teachers across various platforms, each with different educational objectives.\\

One of the major goals of Plan CEIBAL is to improve the English teaching and learning in primary education. In 2021, the Little Bridge platform (\url{https://www.littlebridge.com/}) was incorporated into the teaching of English in primary schools. Little Bridge is a digital platform for learning English that allows students ages 6 to 12 to practice English in an engaging and interactive way. In Uruguay, Little Bridge is used in the last three years of primary school (4th, 5th, and 6th grade).  Additionally, every year, an adaptive English test is administered to a subset of students to evaluate their English proficiency. According to the Common European Framework of Reference for Languages (CEFR), an 11-year-old student should reach the A2.1 level \cite{council2001common}.\\

Learning Management Systems (LMS) or learning platforms like Moodle and Blackboard have become key tools in education. LMS platforms allow users to create, manage, and deliver course content, such as uploading files, grading assignments, or providing students with real-time feedback. A vast amount of student and teacher data is generated by LMS platforms daily. Transforming this data into relevant information for decision-making is a major challenge due to the complexity of the data structure and the difficulty of summarizing the learning process based on it.\\

The data generated from LMS can be used for Learning Analytics (LA) or Academic Analytics (AA) purposes. Learning Analytics (LA) is the “measurement, collection, analysis and reporting of data about learners and their contexts, for purposes of understanding and optimizing learning and the environments in which it occurs” \cite{long2014penetrating} while Academic Analytics (AA), in contrast, is “the application of education analytics for better decision making at institutional, regional, and international levels’’ \cite{long2014penetrating}.\\

This paper focuses on techniques related to Learning Analytics (LA) as it centers on students and how LMS data can be leveraged to understand and enhance student success. Several studies have shown an association between LMS usage and student performance, highlighting the importance of measuring both the quality and intensity of LMS usage for institutions \cite{filippidi2010impact, jo2014analyzing, macfadyen2010mining, whitmer2013logging}. However, research on Learning Analytics has primarily focused on tertiary education due to its strong adoption by universities.

There is significant potential to use Learning Analytics to support students and diagnose their learning progress in a pre-university context \cite{kovanovic2021learning}. As mentioned in \cite{freeman2017nmc}, the main opportunities relate to triggering interventions or curricular adaptations, predicting learner outcomes, and even prescribing new pathways or strategies to improve student success.\\

In \cite{rogers2021developing}, an analysis of Little Bridge data suggests that users who are most active in the platform’s social network also complete more learning activities with better results. However, since the English evaluation is conducted within the Little Bridge platform, it is not possible to determine students’ English performance within the national education system.
This paper presents statistical and computational methods to transform primary education students’ LMS usage data into meaningful insights and uses them to predict student performance. The main objective of this research is to analyze the data generated by the use of the Little Bridge platform in Uruguay and predict students’ performance on the English test.\\

Usage data from students and teachers across various platforms, each with different educational objectives, was provided by Plan CEIBAL, particularly information about Little Bridge use and English test results. Different goals were proposed based on these data, with various data aggregation structures and modifications to the response variable in each case. The problems can be summarized in terms of the predictive variable. For example, it might be relevant to predict the final score obtained by each student in the test (considered as a numerical variable), predict the average score of the class, or predict whether a student will reach the expected level of English by the end of primary school. The results presented here focus on the latter problem, as it is the only one where the educational system specifies an expected English proficiency level.\\

The following specific research questions are proposed to address the main goal of this research:

\begin{itemize}
\item \textbf{Q1.}  Is it possible to estimate the probability of reaching the expected English level based on predictor variables?
\item \textbf{Q2.}  How early can we identify performance issues without negatively affecting the predictive performance of the model?
\item \textbf{Q3.} Can we identify educational centers that differ in performance and learn from them?
\item \textbf{Q4.} Are individual factors, such as platform use, more important than socioeconomic context, class, or educational center in impacting student performance?\end{itemize}
\vspace{0.5cm}

The first question (Q1) concerns the accuracy of a classifier in predicting whether a student will reach the expected level of English. In this work, tree-based methods such as Bayesian Additive Regression Trees (BART) and Random Forest models are used to build flexible predictive models with good performance.

The second question (Q2) addresses the early warning issue. In order to help at-risk students, it is important to identify them as early as possible, but using less information might affect predictive performance. Therefore, it is valuable to study the sensitivity of statistical methods to a time cutoff that defines possible interventions.

The third question (Q3) is related to the fact that education occurs within a social context, making it important to analyze how this context might affect students’ performance. From a statistical perspective, this corresponds to grouping effects in the data, which can be modeled with random effects. These effects help identify the importance of the educational center and potentially flag centers that need support or are performing well, allowing their experiences to be shared with other centers.

The fourth question (Q4) concerns variable importance. Some predictor variables relate to individual factors, such as platform usage and activity results during the school year, while other relevant factors are context variables, such as the educational center and the socioeconomic environment shared by many students in the class. Using the estimated models, it is possible to compare how these two sets of predictors affect the probability of student success.\\

This document is organized as follows: Section II describes the data sources and variables involved in the analysis. Section III describes the statistical methods used in the paper.  Section IV presents the study results, an exploration of the data, and the main results from machine learning models.

Finally, Section V presents a discussion of the main findings and suggests possible directions for future research.

The material needed to reproduce the article's results is available at \url{https://github.com/nachalca/paper_ceibalBART}. However, due to confidentiality reasons, the data must be requested separately.

\section{Data Sources}

The data used for this study was provided by CEIBAL and corresponds to the year 2021, the first year in which the Little Bridge platform was available for students and teachers in primary school.

Two main data sources were used: usage data from Little Bridge (LB), which was the primary source for computing predictor variables, and adaptive English test data, which was mainly used to obtain the response variable.

Table \ref{tab:desc_files} presents the description and size of each file in terms of rows and columns. Due to the large volume of data, the \texttt{data.table} library \cite{data.table} was used in combination with \texttt{dtplyr} \cite{dtplyr} for its simple syntax and computational efficiency. The first stage involved data transformation, where separate files were merged into a single dataset for modeling.

  \begin{table}[hbpt]
    \centering
      \caption{Data files  to combine for 2021}
    \label{tab:desc_files}
    \newcolumntype{M}[1]{>{\centering\arraybackslash}m{#1}}
    \begin{tabular}{|M{4cm}|M{3cm}|}
          \hline
          \textbf{Description} & \textbf{Size} 
          \\
          \hline
           Student activity in Little Bridge. & 1030280 rows, 19 columns.
          \\
          \hline
      Logs of received and sent messages. & 585082 rows, 15 columns.         
          \\
          \hline
          Lessons assigned by teachers to the class.  &  84449 rows, 7 columns.
          \\
          \hline
           Student's Adaptive English test answers. &  35032 rows, 13 columns.
          \\
          \hline
          Student's Adaptive English Reading test answers & 33309 rows, 13 columns.
          \\
          \hline
         Activity name and  the lesson it belongs to & 915 rows, 3 columns.
          \\
          \hline
    \end{tabular}
  \end{table}
  
\begin{figure}[hbpt]
\includegraphics[width=\columnwidth]{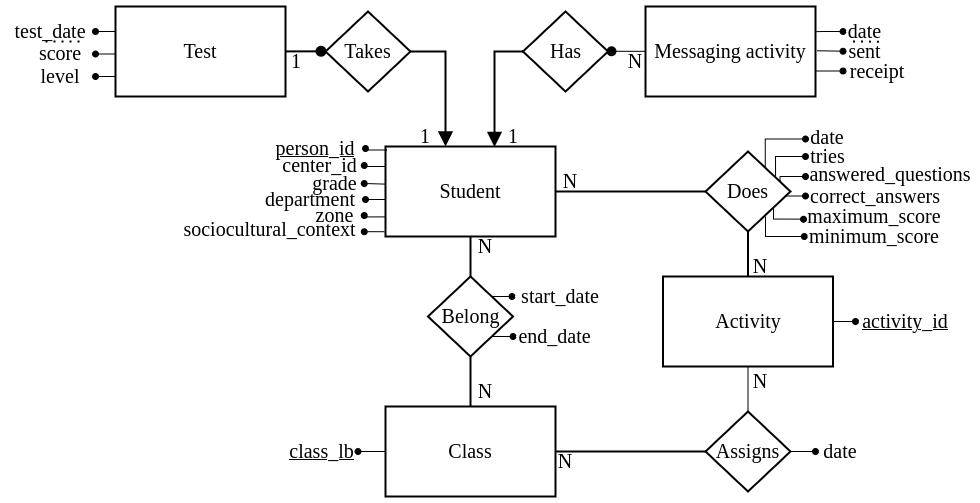}
\caption{ Entity-relationship diagram}
\label{ER}

\end{figure}

To facilitate data interpretation and establish a foundation for further development, the data was represented using an entity-relationship model. Fig. \ref{ER} presents the corresponding entity-relationship diagram \cite{chen1976entity}.

To construct this representation, data cleaning and transformation were performed, requiring the adoption of specific criteria. These decisions were essential to ensure the data could be effectively used in a predictive model. The following section outlines the criteria applied to represent each entity.\\

\textbf{Students:} Each student is characterized by an identifier, along with their school, grade, department, zone, and socio-cultural context. Although these variables may change over time, the provided data does not reflect such changes. Students who were not assigned to any class, did not complete any activities, or whose activities were not recorded in the dataset provided by CEIBAL were filtered out. This filtering was applied because the objective of this study is to link platform usage with the final outcome obtained.\\

\textbf{Test:} The adaptive English test is taken once a year at the end of the school term. The test results are represented by the date it was taken, the score obtained, and the level. The highest recorded score was 900, while the lowest was 225.21. The level is determined based on the score using predefined cutoff points.
Table \ref{tab:niveles} shows the cutoff points for each test level.
In the original dataset, some students had multiple test records. In these cases, only the result from the first test taken was selected. If two attempts were made on the same day, the attempt with the lowest score was chosen.

  \begin{table}[htbp]
    \centering
      \begin{tabular}{|c|c|}
      \hline
      Level & Points \\
      \hline
      Pre-A.1 & $< 372.9$ \\
      \hline
       A1.1 & $372.9 - 444.4$ \\
      \hline
       A1.2 & $444.4 - 495.5$ \\
      \hline
       A2.1 & $495.5 - 527.9$ \\
      \hline
       A2.2 & $527.9 - 599.4$ \\
      \hline
      B1 & $> 599.4$ \\
      \hline
      \end{tabular}%
    \caption{English test cut points for each level}
    \label{tab:niveles}%
  \end{table}%

\textbf{Message activity:} The message activity entity is defined by the date, the number of messages sent and received, and the number of threads. Only days when a message was sent or received are recorded. Very few actions were required; specifically, null values were replaced with 0, and March 1, 2021, was designated as the activity date for these cases. Message activity is determined by the student associated with it and the date on which it occurred.\\

\textbf{Activity:} The activity entity is defined solely by its identifier. Activities are part of lessons, but since there is no information available regarding the difficulty of the lessons or any other relevant details, they were excluded from the model. Teachers assign lessons to the class. Since an activity can be associated with more than one lesson, only the first assignment is recorded. It is important to note that the assignment by the teacher is made at the class level, not the student level. For each activity, the number of times it was completed, the questions answered, the number of correct answers, and the maximum and minimum scores across attempts are registered. If a student only attempted the activity once, the maximum and minimum scores are the same.\\

\textbf{Class:} A class is represented by its identifier. Students can change classes over time. Since the exact date a student was enrolled in a new class is not available, the enrollment date is considered the first activity record for the student in that class, and the unenrollment date is the day before. A student cannot belong to two classes simultaneously during the same period. Based on the proposed entity-relationship model, we generated three final datasets: student data, sixth-year student data, and class data. The first and second datasets use students as the unit of analysis, while the third dataset uses classes as the unit of analysis.\\

Table \ref{tab:variables} provides a brief description of the 161 variables used in the predictive models. Many of the predictor variables are designed to summarize student activity on the LB platform, both on a monthly basis and cumulatively throughout the year. Consequently, in many cases, each row in the table corresponds to 9 variables—one for each month from March to November. Examples of such variables include cumulative activities up to a specific month mm (cum\_act\_mm), average accuracy in month mm (acc\_mm), cumulative correct answers up to month mm (cum\_correct\_mm), sociocultural context, among others.



\section{Statistical Methods} \label{sec:stat}
This section describes the statistical models used to predict whether students reach the expected English level.

Let $ D_n = \{(\mathbf{x}_i, y_i)\}_{i=1}^n $ denote the training dataset, where $ \mathbf{x}_i \in \mathbf{R}^p $ is a vector of explanatory or predictor variables for oservation $i$, and $y_i$ the response variable. 

\subsection{Bayesian Additive Regression Trees (BART)}
Bayesian Additive Regression Trees (BART) is a non-parametric machine learning method introduced in \cite{chipman2010bart} that provides both high predictive accuracy and strong interpretability, particularly in causal inference settings \cite{hill2011bart}. BART is applicable to both continuous and binary outcomes, and can model hierarchical structures by incorporating random effects. As a Bayesian method, BART enables uncertainty quantification via posterior distributions. Extensions for multinomial, survival, and other outcome types are reviewed in \cite{hill2020bartreview}, additional flexibility in the model has recently being introduced in \cite{linero2024_generalizedBART}. 

\subsubsection{BART data model}
As a Bayesian model, a probabilistic model for data generation is assumed. In particular, the BART model assumes the following form:
$$
y_i = f(\mathbf{x}_i) + \epsilon_i
$$
where $y_i \in R$ and $\epsilon_i \sim \mathcal{N}(0, \sigma^2)$. When the response variable is binary, BART model can be modify using a probit link:
$$
\Pr(y_i = 1 | \mathbf{x}_i) = \Phi( f(\mathbf{x}_i) )
$$
where $\Phi()$ is the cumulative distribution function of the standard normal distribution. 

The main goal is to obtain an estimation for $f()$, BART uses an ensemble of decision trees constructed using a \textit{boosting} strategy; that is, the function $f()$ is modeled as a sum of regression trees:
$$
\begin{array}{rcl}
f(\mathbf{x}_i) & = & \sum_{b=1}^{B} g(\mathbf{x}_i; \mu_b, T_b) \\
g(\mathbf{x}_i; \mu_b, T_b) & = &\sum_{j=1}^{J_b} \mu_{bj} I(x \in T_{bj})
\end{array}
$$
where $T_b$ represents a tree structure, describe as a sequences of decision rules that partition the feature space into $\left\{ (T_{bj} \right\}_{b=1}^{J_b}$ regions, and $\mu_{b} \in R^{J_b}$ is a vector with values $\mu_{bj}$ associated to each $T_{bj}$ region.  In short, tree $g(\mathbf{x}_i; \mu_b, T_b)$ partitions the feature space into regions and assigns a constant prediction $\mu_{bj}$ within each region. 

\subsubsection{Inference for BART model}
Statistical inference in Bayesian settings requires to specify a prior distribution for all model parameters and usually requires a Markov Chain Monte Carlo (MCMC) sampling algorithm to approximate the posterior distribution. Here, we outline the prior and the sampling scheme for a basic version of the algorithm; a detailed description of both aspects can be found in \cite{bartmachine}. 

Priors distributions are specified for the tree structures $T_b$, the terminal node parameters $\mu_b$, and noise standard deviation, $\sigma$.  The joint prior is structured as follows:
$$
p(T_b, \mu_b, \sigma) = p(\sigma) \prod_b p(T_b) \left ( \prod_j p(\mu_{bj}|T_b) \right)
$$
this implies that \textit{a priori} $\sigma$ is independent of all trees and each tree's parameters are mutually independent. The prior for the tree structure is set up so that the probability of split a particular node at depth $d$ is proportional to 
$$
\alpha(1+d)^{-\beta}
$$
where $\alpha \in (0,1)$ and $\beta > 0$. Conditional on the tree structure, the rest of the priors are set up to have conditional conjugacy, 
$$
\begin{array}{cc}
\mu_{jb}|T_b \sim N(0, \sigma_{\mu}^2) &  \sigma^2 \sim \chi^{-2}(\lambda)  
\end{array}
$$
where $\sigma_{\mu}^2$ is set so the posterior expected value lies in the range of data with 95\% probability, and $\lambda$ is set so $\sigma^2$ is lower than the observed residual variance of a linear regression with 95\% probability.  

The posterior distribution $p(T_b, \mu_b, \sigma^2|D_n)$ is approximated via simulation using a Gibbs sampling algorithm based on \textit{Bayesian backfitting} (\cite{hastie2000bayesian}),  where, at each iteration, an individual tree is fitted conditional on the rest of the trees. This is done by using the residuals $R$ and $R_{-b}$ defined as: 
$$
\begin{array}{ll}
    R &= y - \sum_{b=1}^B g(\mathbf{x}_i; \mu_b, T_b) \\
    R_{-b} &= y - \sum_{j\neq b} g(\mathbf{x}_i; \mu_j, T_j) 
\end{array}
$$
The Gibbs algorithm is outlined in \ref{algo.gibbs}, each draw is obtained from a full-conditional distribution. Tree structure, $T_b^{(i)}$ is sampled using a Metropolis-Hasting step, where a modified version of the current tree is proposed and then randomly accepted (or not). The cases for $\mu_b^{(i)}$ and $\sigma^{2^{(i)}}$ are simpler due to their conjugacy properties.

\begin{algorithm} 
\KwIn{Previous samples values $(T_b, \mu_b, \sigma^2)^{(i-1)}$ and $R^{(i-1)}$ }
\KwOut{New samples $(T_b, \mu_b, \sigma^2)^{(i)}$ and $R^{(i)}$ }

\For{$b = 1$ \KwTo $B$}{
    Compute $R_{-b}^{(i-1)}$
    Draw $T_b^{(i)}$ from $p\left(T \mid R^{(i-1)}_{-b}, \sigma^{2^{(i-1)}}\right)$\;
    Draw $\mu_b^{(i)}$ from $p\left(\mu \mid T_b^{(i)}, R^{(i-1)}_{-b}, \sigma^{2^{(i-1)}}\right)$\;
}
Draw $\sigma^{2^{(i)}}$ from $p\left(\sigma^2 \mid (T_1, \mu_1), \ldots, (T_B, \mu_B), R\right)$\;
Compute $R^{(i)}$;
\caption{Gibbs sampler \label{algo.gibbs} }
\end{algorithm}

\subsubsection{Clustered data}
Educational data often exhibit a hierarchical or clustered structure (e.g., students within classes, and classes within schools). These groupings can introduce intra-group correlations, which are often substantively meaningful and should be accounted for in the analysis. 

Cluster effects can be incorporated in BART model as a random intercept (\cite{carnegie2019examining}, \cite{tan2018_bart}) for each group:
$$
y_{ig} = f(\mathbf{x}_i) + u_g + \epsilon_{ig}
$$
where $u_g \sim N(0, \sigma_u^2)$ is a group-level random effect. Then, observations from the same group are no longer independent since they share the $u_g$ term, additionally this approach allows to isolate the impact of the school context on student outcomes.

\subsection{Random Forest}

\textit{Random Forest} (RF) is a supervised ensemble learning method that builds multiple decision trees and combines their predictions to improve accuracy and reduce overfitting \cite{breiman2001random}. This ensemble method provides strong predictive accuracy and is often used as a benchmark in classification tasks. In this study, we compare Random Forest with BART in terms of predictive performance, while also highlighting BART’s advantages for interpretability and multilevel modeling.

RF can be applied to both classification and regression problems. Here, we describe its use for classification tasks. That is, $y_i \in \mathscr{G} $ represents a class label for observation $i$. Let $\mathscr{G} = \{1, 2, \dots, G\}$ denote the set of possible class labels.

Random Forest consists of an ensemble of decision trees $\{h(\mathbf{x}, \Theta_k)\}_{k=1}^B $, where $h(\cdot)$ denotes a randomized classification tree. RF uses two key techniques to increase model diversity and reduce variance: (1) bootstrap sampling of the training data, and (2) random feature selection at each tree node. These strategies promote the creation of de-correlated trees, whose aggregated predictions are more robust than any individual model. In the randomized tree $h(\mathbf{x}, \Theta_k)$, $\Theta_k $ is a random vector controlling the bootstrap sample and feature selection for the $ k $-th tree.

The final prediction is determined by majority vote across all trees. The theoretical prediction function is given by:
\begin{equation} \label{rfesti}
f(\mathbf{x}) = \operatorname*{arg\,max}_{g \in \mathscr{G}} \mathbb{P}_{\Theta}(h(\mathbf{x}, \Theta) = g)
\end{equation}

In practice, the probability is approximated by averaging over $ B $ bootstrap samples:
\begin{equation} \label{predfor}
\hat{f}(\mathbf{x}_0) = \operatorname*{arg\,max}_{g \in \mathscr{G}} \sum_{k=1}^B I(h(\mathbf{x}_0, \Theta_k) = g)
\end{equation}
where $ I(\cdot) $ is the indicator function.

An outline of the RF algorithm to obtain $\hat{f}(\mathbf{x}_0)$ is shown in \ref{rf.algo}


\begin{algorithm}[H]
\DontPrintSemicolon
\KwIn{$d_n$, $B$,  $p$, $m$}
\KwOut{RF: $\{h(\mathbf{x}, \Theta_k)\}_{b=1}^B$}

\For{$b = 1$ \KwTo $B$}{
    Draw a bootstrap sample $b_k$ from $d_n$\;
    Use $b_k$ to grow $h(\mathbf{x}, \Theta_k)$:\;
    \Indp
    All cases in $b_k$ at the root node\;
    \While{stopping criterion not met}{
        Randomly select $m \ll p$ predictors\;
        Choose the best split among the $m$ variables\;
        Split the node into two child nodes\;
    }
    \Indm
}
The final model is an ensemble of trees: $\{h(\mathbf{x}, \Theta_k)\}_{b=1}^B$\;
\caption{Random Forest construction \label{rf.algo}}
\end{algorithm}

\section{Results}
This section presents the main results of the data analysis. The first subsection provides several graphical summaries of the dataset to gain insights into the statistical problem before applying any machine learning methods. Then, the main results of the statistical learning approaches are presented, guided by the specific research goals outlined in the Introduction.

\subsection{Implementation details}
BART is implemented using the \texttt{dbarts} package \cite{dbarts}, while RF is implemented with the \texttt{ranger} package \cite{ranger}. Several hyperparameters need to be choose to fit both BART and RF models. Hyperparameters in BART corresponds to the values in the prior distributions, and the number of trees, and number of iterations and chains for the MCMC algorithm. Hyperparameter in RF corresponds to the number of trees, maximum depth of individual trees, number of variables considered for each split.

Optimal hyperparameters for each model are determined using different techniques. For BART, we employ the \texttt{ParBayesianOptimization} package \cite{ParBayesianOptimization}, which automates hyperparameter tuning based on the Bayesian Optimization framework (implementing methods described in \cite{snoek2012practical}). For RF, we use the \texttt{tidymodels} framework \cite{tidymodels}, specifically the \texttt{dials} package \cite{dials}, which provides functions for Bayesian optimization of hyperparameters.

\subsection{Data set exploration}
\begin{figure}[hbpt]
 \includegraphics[trim=0 4cm 0 4cm, clip, width=\columnwidth]{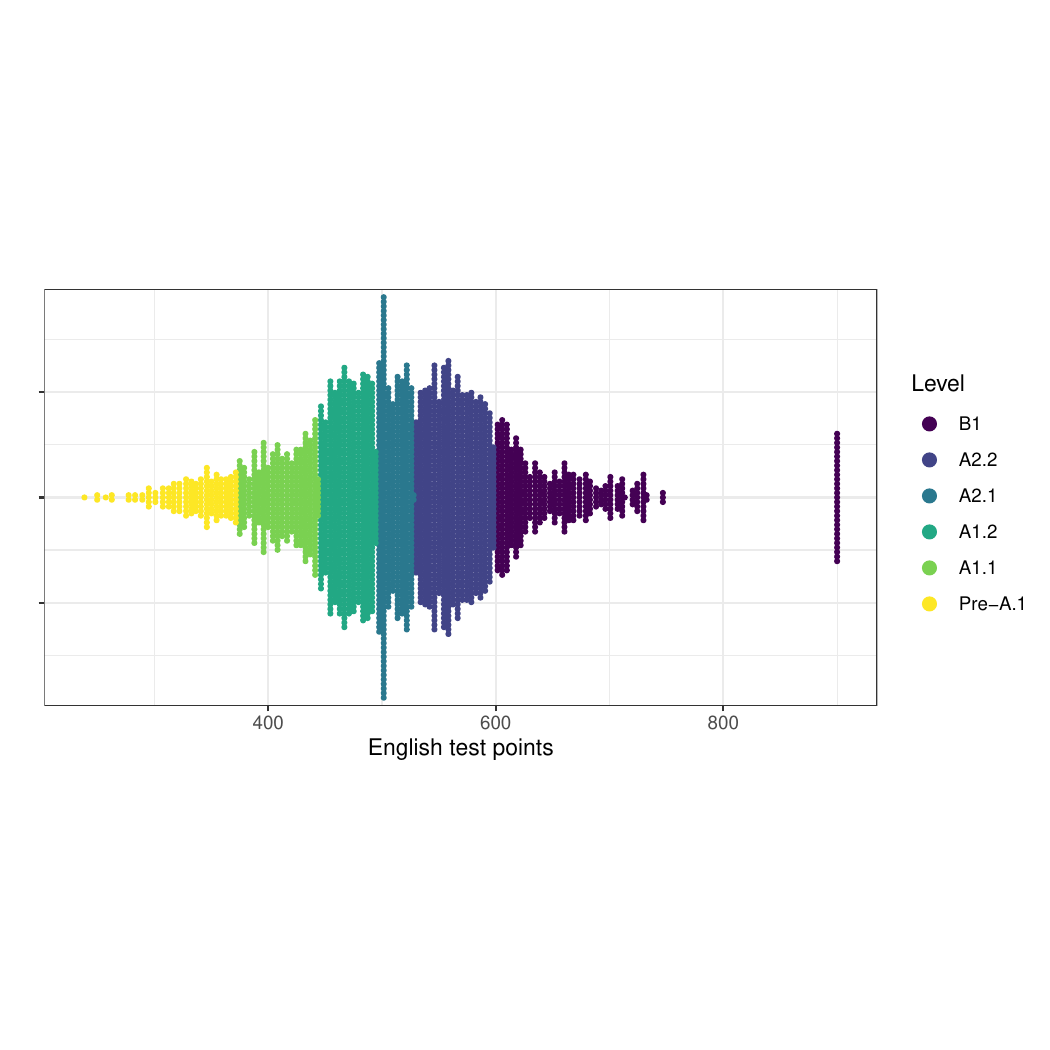}
\caption{Dotplot plot of 6th-grade students' English test results. Colors represent the test levels.}
\label{sina}
\end{figure}

In the Uruguayan educational system, sixth grade is the final year of primary school. In a national English test, students are expected to achieve at least the A2.1 English proficiency level, with proficiency levels defined in Table \ref{tab:niveles}. Figure \ref{sina} displays a sina plot of English test results for sixth-grade students, with colors indicating the performance level attained. The plot reveals that a large proportion of students—approximately 40\%—do not meet the expected A2.1 benchmark.

\begin{figure}[hbpt]
\includegraphics[width=\columnwidth, height=0.7\columnwidth]{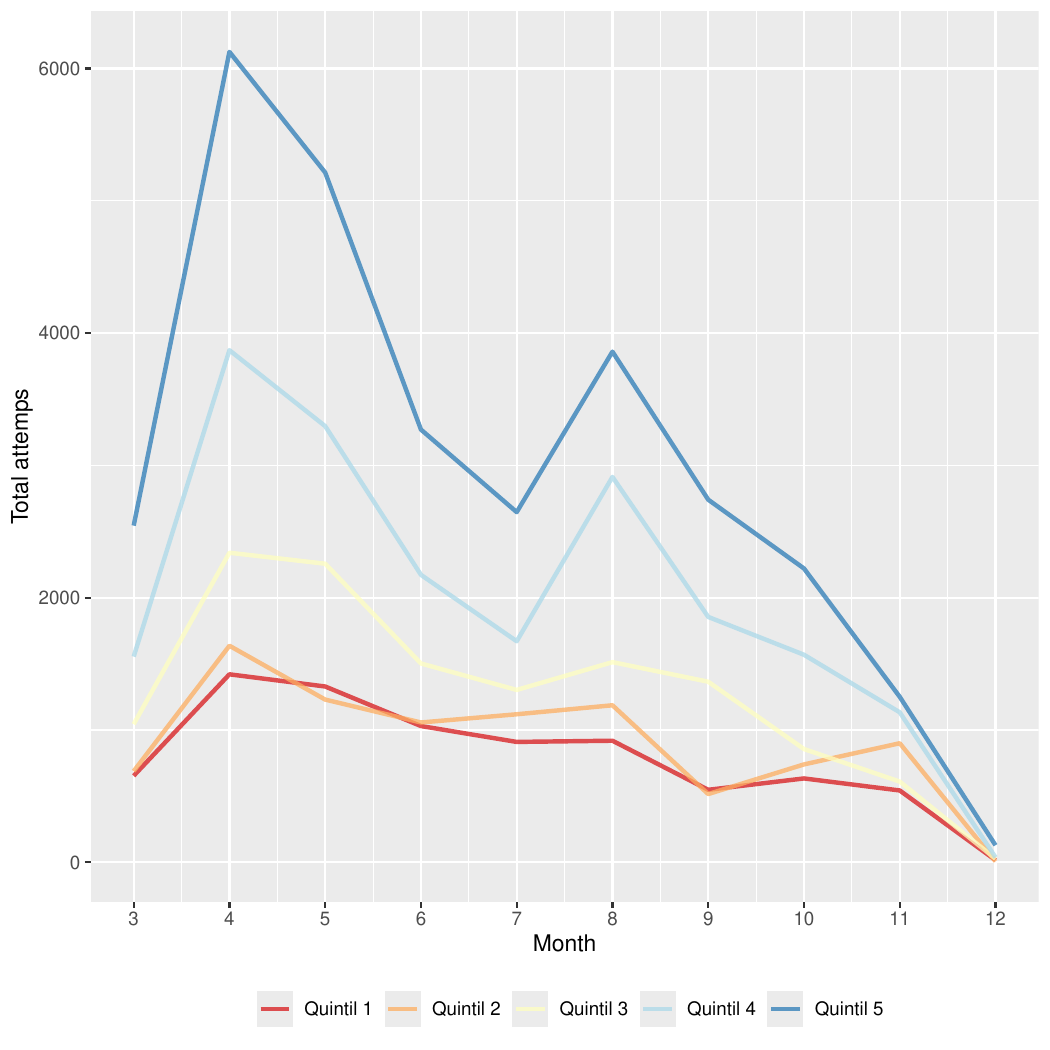}
\caption{ Monthly Total LB activity attempts over all students by socioeconomic group (represented by color). }
\label{att}
\end{figure}

An initial summary of students' usage of the LB platform can be obtained by aggregating the total number of activities attempted by each student. Figure \ref{att} shows the total number of activity attempts per month, with school socioeconomic context represented by color. The color patterns indicate that students from schools with higher socioeconomic status tend to have more activity attempts. The lowest levels of participation are consistently observed in Quintiles 1 and 2. Additionally, in terms of temporal patterns, April and August stand out as the months with the highest activity levels.

\begin{figure}[hbpt]
\includegraphics[trim=0 3cm 0 3cm, clip, scale=.8]{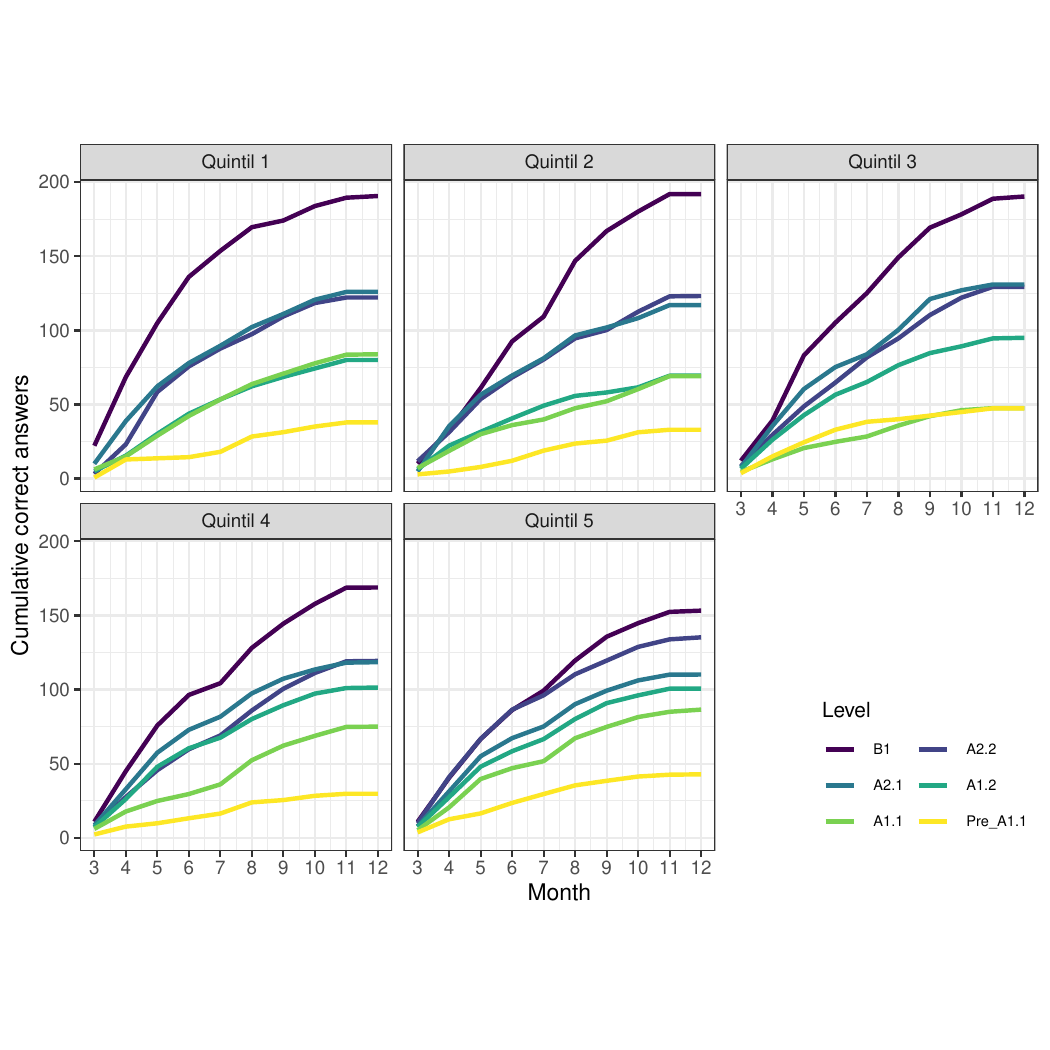}
\caption{ Cumulative correct answers total by month over all students by test level group (represented by color) in each socioeconomic group (represented by facet panels).}
\label{useperf}
\end{figure}

Finally, is interesting to explore the relationship between LB platform activity performance and English test results. Figure \ref{useperf} shows the total of correct answers in every school for each level of socioeconomic context by month. All socioeconomic groups show a positive relationship between performance on LB (measured by correct answers) and performance on the English test, also that LB performance separate the test levels groups after a few months indicating these data could be use as early warnings. 


\subsection{Model results}

This subsection presents the main modeling results, addressing the research questions posed in Section \ref{sec:Intro}.

\textbf{Predictive Modeling and Early Warning (Q1 and Q2)} The first two questions, relate to fitting a suitable model for predicting student success. A classification problem is formulated using the response variable $Y_{ig}$, defined as 
$$
Y_{ig} =\begin{cases}
       1  & \text{student } i \text{ in group } g \text{ reaches A2.1 level or higher} \\
       0  & \text{ otherwise }
    \end{cases}
$$
The predictor variables, denoted as $X_{ig}^{m}$, represent all available information on LB usage for a student, accumulated up to month $m$. The time frame for this information ranges from the beginning of the school year in March ($m=3$) to November ($m=11$), when the English test is administered. For each time point, four predictive models are trained: RF and BART models, each with both default and optimized hyperparameters. 

\begin{figure}[hbpt]
      \centering
      \includegraphics[width=\columnwidth]{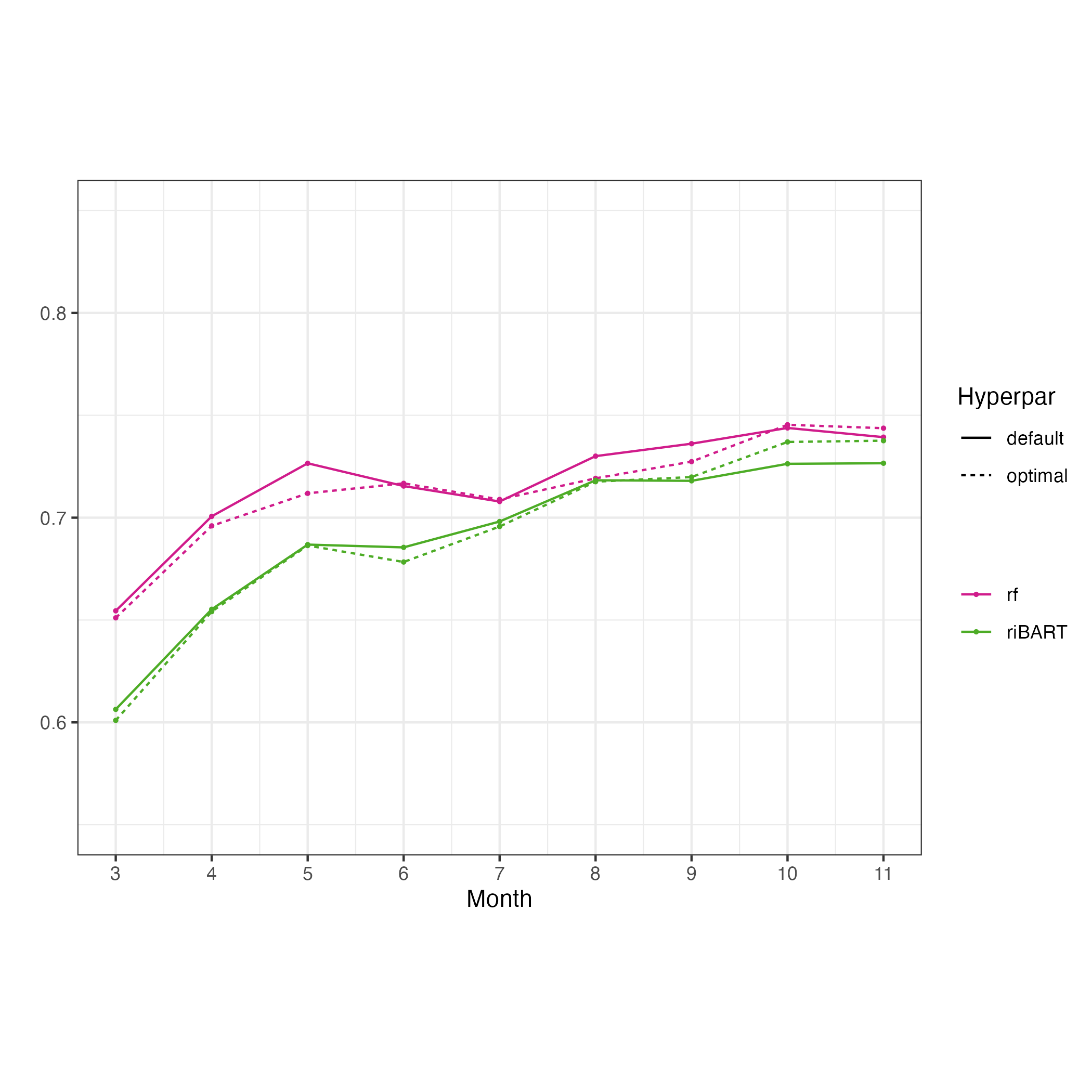}
      \caption{AUC results on the test dataset. The horizontal axis represents the month of the data collection cut-off point, while the vertical axis shows the AUC performance measure. The color indicates the predictor model, with dashed lines representing tuned hyperparameters and solid lines indicating default hyperparameter values.}
      \label{fig:auc}
\end{figure}

Area under the Curve (AUC) performance measure, as a summary of the Receiving Operating Characteritic (ROC) curve, is a widely used performance metric for classification problems, Figure \ref{fig:auc} shows AUC for each model at each time point. As expected, using more months of data improves model performance: AUC increases from 0.60–0.64 (March data) to above 0.70 with full-year data. However, most of this gain occurs in the early months, suggesting the possibility of identifying at-risk students as early as May with acceptable accuracy. In what follows, all the results are obtained using setting $m=5$ as the cut point limit.

Table \ref{perftab} compares predictive performance of Random Forest and BART models, both with optimal tuning and using data up to May ($ m = 5 $). Although Random Forest performs slightly better, BART is chosen for further analysis due to its interpretability, uncertainty quantification, and ability to include random effects.

\begin{table}[!t] 
\caption{Performance measures with data up to May} 
\label{perftab}
\centering
\begin{tabular}{l|cc}
\hline
Measure & RF & BART \\ \hline
accuracy & 0.67 & 0.65 \\ 
sens & 0.76 & 0.76 \\ 
spec & 0.53 & 0.47 \\ 
roc\_auc & 0.71 & 0.69 \\ 
\hline
\end{tabular}
\end{table}

\textbf{School Effects via Random Intercepts (Q3)} In order to assess the impact of school context, we extract school-level random intercepts from the BART model. A positive value suggests the school performs better than expected given its student profile; a negative value indicates underperformance.  Posterior samples for each center effect are obtained and used to construct a point estimate and its variability.

\begin{figure}[hbpt]
      \centering
      \includegraphics[width=\columnwidth]{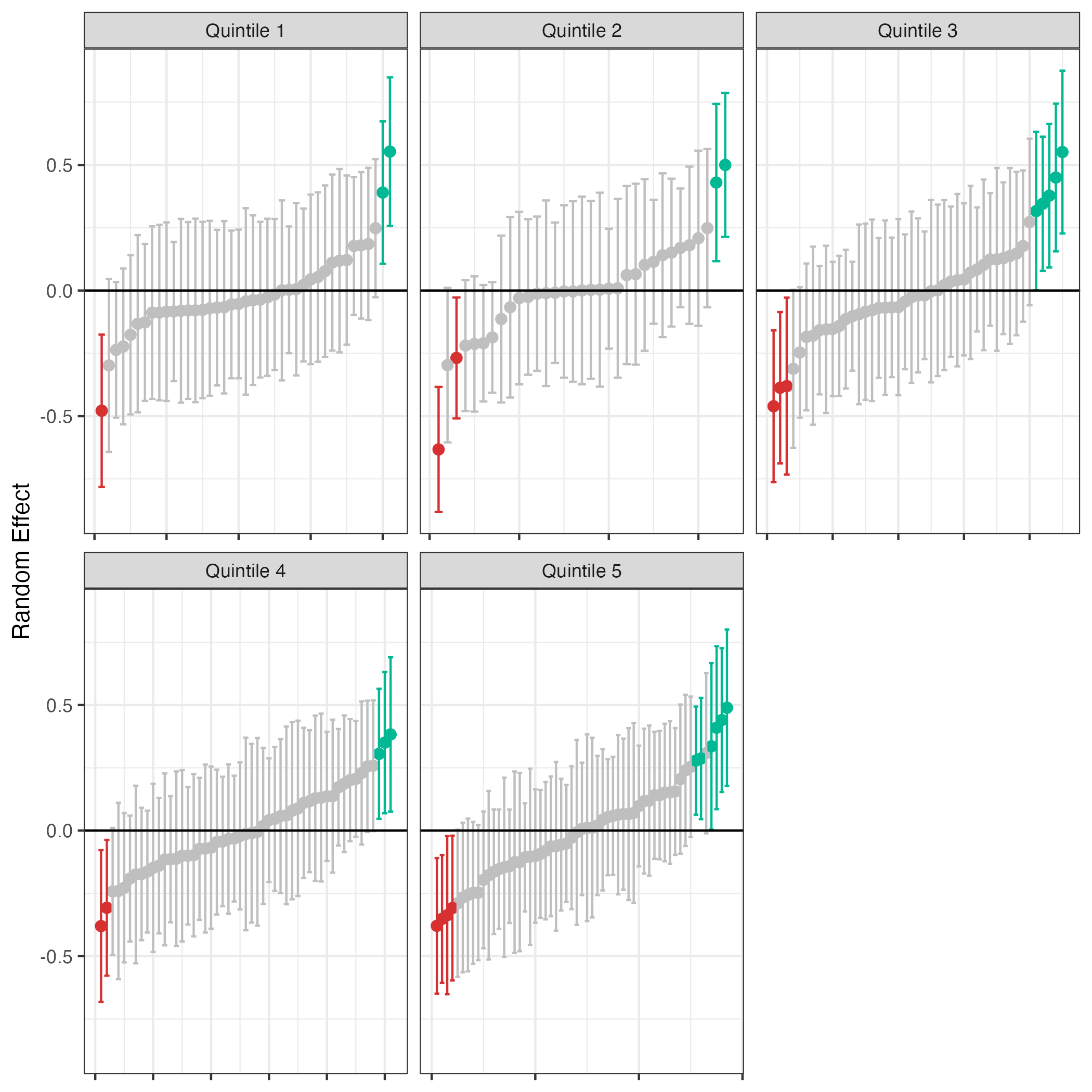}
      \caption{Random effects of schools by socioeconomic quintile. Schools with positive and negative random effects are highlighted. The error bar represents one standard deviation.}
      \label{randomeff}
\end{figure}

Figure \ref{randomeff} presents estimated school effects by socioeconomic quintile, along with one standard deviation error bars. Schools with credible positive or negative effects (error bars not overlapping zero) are highlighted. Twelve schools show significantly negative effects, while sixteen display significantly positive effects. High-performing schools appear across all socioeconomic levels, indicating that effective practices may be transferable. Similarly, identifying underperforming schools enables targeted interventions.

\textbf{Individual Platform Usage vs Socioeconomic Context (Q4).} Finally, the last question concerns understanding the impact of LB platform usage compared to the influence of student context, as reflected in the socio-economic context of the educational center. It's important to note that this is not a trivial task, as LB usage is represented by multiple predictor variables, making it challenging to characterize its effect using traditional tools, such as partial dependence plots.

Three synthetic students, each representing a different LB platform usage profile, were defined. A low usage profile was created, where all predictor variables related to LB usage are set at the observed 10th percentile. Similarly, an average usage profile was constructed with usage variables at the 50th percentile, and a high usage profile with variables at the 90th percentile. These three profiles were repeated across all socioeconomic levels, resulting in 15 synthetic students in total. Other attributes unrelated to platform activity were kept constant: all individuals were assigned to the same city (Montevideo), classified as urban, and placed in the same class.

\begin{figure}[hbpt]
      \centering
      \includegraphics[width=0.8\linewidth]{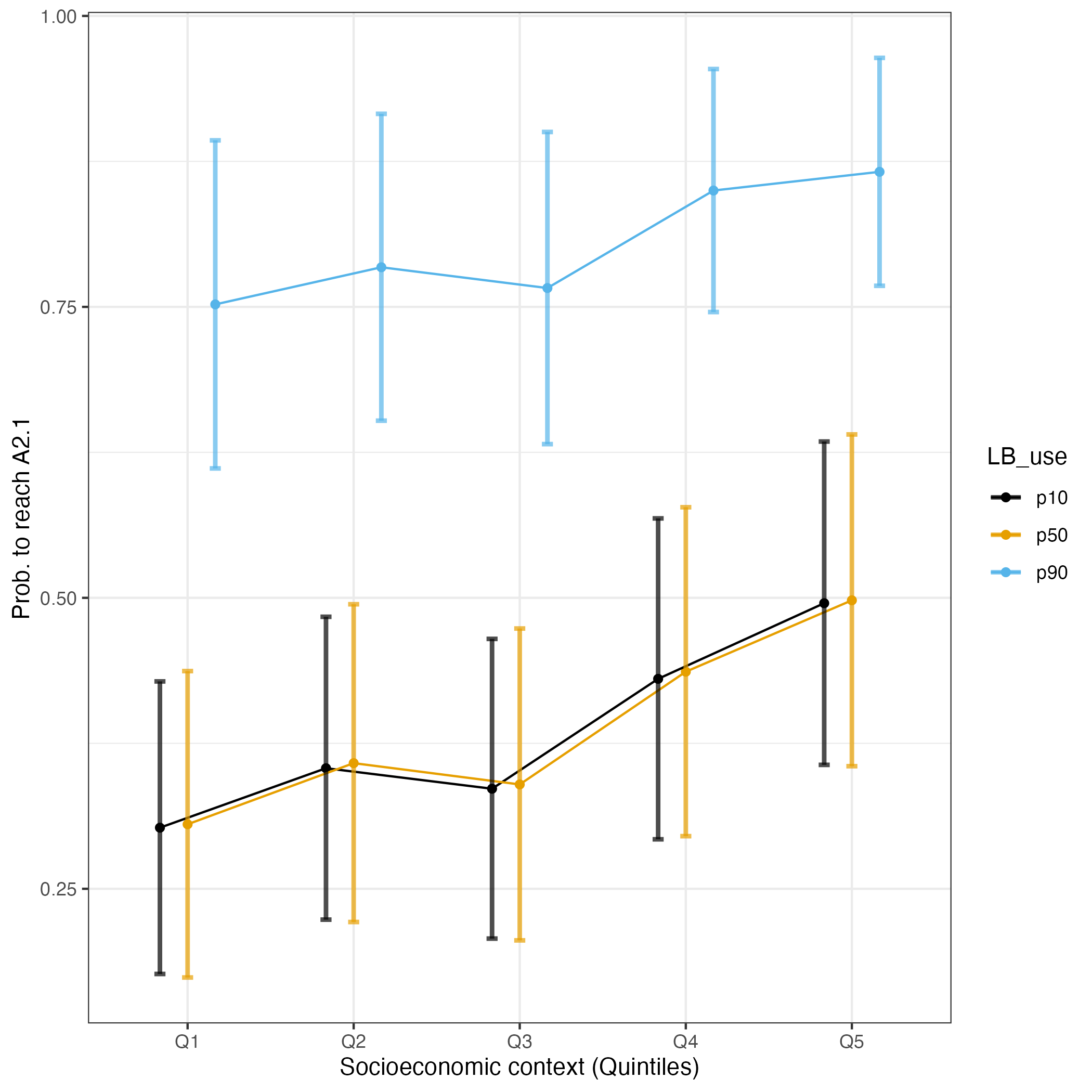}
      \caption{Probability of reaching the expected English level by socioeconomic context and LB usage intensity for synthetic students.}
      \label{synte}
\end{figure}

The results are presented in Fig. \ref{synte}, where the y-axis represents the probability of reaching the expected English level (A2.1), the x-axis corresponds to socioeconomic quantiles, and the color indicates the level of LB usage intensity.  We observe that students with high LB usage (90th percentile) are more likely to reach the expected English level, with this probability increasing across socioeconomic quantiles. When comparing a student with intense usage in a lower socioeconomic context (Quantile 1) to a student with lower usage in a higher socioeconomic context (Quantile 5), we find that platform usage has a greater impact than socioeconomic context in achieving the expected English level.

\section{Discussion and Conclusion}
LMS data are leveraged to extract relevant insights for the educational system using appropriate computational tools and statistical methods. Machine learning techniques are employed to construct predictive models for student academic performance.

Predictive performance, measured by AUC, ranges from 0.6 to over 0.7 with varying data lengths. Most improvement occurs within the first 3 months. These models can serve as early-warning tools to identify students at risk. The BART model incorporates a random intercept to account for school effects, which can inform school-specific strategies. The BART model also examined the relationship between LB platform usage and school socioeconomic context on the English test. Synthetic data showed that high LB usage (90th percentile) increases the probability of reaching the expected English level across socioeconomic quintiles. Notably, platform usage has a greater impact than socioeconomic context on reaching the expected English level.

Our findings offer actionable insights for educators and decision-makers. First, the ability to accurately identify at-risk students early enables teachers to implement timely and targeted interventions, underscoring the value of integrating predictive analytics into ongoing assessment cycles. Second, the model’s capacity to distinguish high-performing schools across all socioeconomic levels provides a pathway to identifying effective pedagogical practices that can be transferred to other contexts. Lastly, the stronger predictive value of platform usage relative to school context suggests that digital engagement is a promising lever for promoting equity in learning outcomes. Overall, this study demonstrates how statistical learning tools can transform raw LMS data into a practical framework for supporting equitable, data-informed teaching strategies in primary education.

This study has limitations. The data are observational, prone to noise from classroom computer tool usage. Some children may receive English outside elementary school, reducing the link between LB training and test results. The grouped data structure lacks class or teacher effects, the available R packages do not support fitting multiple effects simultaneously. However, a recent implementation of the BART model in STAN software can accommodate more flexible modeling structures (see \cite{dorie2022stan}). Future research should confirm these findings in an experimental setting with more control over LB usage and its relationship to test results. Using STAN+BART to fit more complex grouped structures, isolating school and teacher effects, is also an interesting next step.

\section{Appendix}

 \begin{table}[htbp]
 \caption{Variable description. Variables ending with mm are repeated for month from 3 to 11} \label{tab:variables}%
    \centering
     \begin{tabular}{|p{3.1cm}|p{4.5cm}|}  
      \hline
      \textbf{Variable}& \textbf{Description}  \\ \hline
id\_person              & Unique student identifier\\ \hline
department              & City in Uruguay, 19 levels\\    \hline
zone                    & Rural or Urban\\                  \hline
sociocultural\_context  & Sociocultural context, 5 levels\\\hline
id\_center              & Student's school id\\ \hline
points                  & English adaptive test scores\\ \hline
cum\_act\_mm         & Cumulative activities up to month mm (from 3 to 11)     \\  \hline
cum\_qs\_mm          & Cumulative questions asked up to month mm (from 3 to 11) \\ \hline
cum\_correct\_mm     & Cumulative correct answers up to month mm (from 3 to 11) \\ \hline
att.prom\_mm         & Ratio of attempts made to activities completed in month mm (from 3 to 11)\\ \hline
activ.prom\_mm       & Average activities completed per day in month mm (from 3 to 11)\\ \hline
qs.prom\_mm          & Average number of questions per activity month mm (from 3 to 11) \\ \hline
pts.min\_mm          & Average minimum points per activity in month mm (from 3 to 11)\\ \hline
pts.max\_mm          & Average maximum points per activity in month mm (from 3 to 11)\\ \hline
acc\_mm              & Average accuracy in month mm (from 3 to 11)\\ \hline
cum\_days\_mm        & Number of days on which activities were performed month mm (from 3 to 11) \\ \hline
cum\_attempts\_mm    & Cumulative attempts up to month mm (from 3 to 11) \\ \hline
cum\_assign\_act\_mm & Cumulative assigned activities up to month mm (from 3 to 11) \\ \hline
perc\_act\_done\_mm  & Ratio between cum\_act\_mm and cum\_assign\_act\_mm \\ \hline
cum\_mess\_send\_mm  & Cumulative messages sent up to month mm (from 3 to 11)\\ \hline
cum\_mess\_rec\_mm   & Cumulative messages received up to month mm (from 3 to 11)\\ \hline
cum\_mess\_thr\_mm   & Number of threads up to month mm (from 3 to 11)\\ \hline
class\_lb\_mm        & LB class to which the student belongs to month mm (from 3 to 11) \\ \hline
reachA2.1                & Response variable, Reach A2.1 level or not\\ 
     \hline
      \end{tabular}%
  \end{table}%
  
\newpage

\section{Acknowledgment}

This paper is part of the research project Sectoral Fund for Digital Inclusion: Education with New Horizons 2020 – Modality A (ref. FSED\_2\_2020\_1\_163528), funded by ANII (\url{https://www.anii.org.uy}) in collaboration with Fundación Ceibal (\url{https://fundacionceibal.edu.uy}).



\bibliography{biblio}

\end{document}